\documentclass[conference]{IEEEtran}
\usepackage{amsmath,amssymb,amsfonts}
\usepackage{algorithmic}
\usepackage{graphicx}
\usepackage{textcomp}
\usepackage{xcolor}
\usepackage{times}
\usepackage{enumerate}
\usepackage{amsmath,amssymb,amsfonts}
\usepackage{mathtools} 
\usepackage{amsthm} 

\usepackage{cleveref}
\usepackage{mathrsfs}
\usepackage{comment}
\usepackage{cite}

\def\BibTeX{{\rm B\kern-.05em{\sc i\kern-.025em b}\kern-.08em
    T\kern-.1667em\lower.7ex\hbox{E}\kern-.125emX}}

\usepackage{placeins}
\usepackage{amsthm}
\newtheorem{definition}{Definition}
\newtheorem{proposition}{Proposition}
\newtheorem{theorem}{Theorem}
\newtheorem{corollary}{Corollary}
\newtheorem{remark}{Remark}
\newtheorem{lemma}{Lemma}
\ifCLASSINFOpdf
\else
\fi
\begin{document}
%
\title{Generalizing the Fano inequality further}
%
%
%

\author{
\IEEEauthorblockN{Raghav Bongole, Tobias J. Oechtering, and Mikael Skoglund}\\
\IEEEauthorblockA{Department of Information Science and Engineering (ISE)\\
KTH Royal Institute of Technology\\
}
\thanks{J. Doe and J. Doe are with Anonymous University.}
\thanks{Manuscript received April 19, 2005; revised September 17, 2014.}}

%
%

\markboth{Journal of \LaTeX\ Class Files,~Vol.~13, No.~9, September~2014}%
{Shell \MakeLowercase{\textit{et al.}}: Bare Demo of IEEEtran.cls for Journals}
%



\maketitle


{\begin{abstract}
Interactive statistical decision making (ISDM) features algorithm-dependent data generated through interaction.
Existing information-theoretic lower bounds in ISDM largely target expected risk, while tail-sensitive objectives are less developed.
We generalize the interactive Fano framework of Chen et al. by replacing the hard success event with a randomized one-bit statistic representing an arbitrary bounded transform of the loss.
This yields a Bernoulli \(f\)-divergence inequality, which we invert to obtain a two-sided interval for the transform, recovering the previous result as a special case.
Instantiating the transform with a bounded hinge and using the Rockafellar--Uryasev representation, we derive lower bounds on the prior-predictive (Bayesian) \(\textup{CVaR}_\alpha\) of bounded losses.
For KL divergence with the mixture reference distribution, the bound becomes explicit in terms of mutual information via Pinsker's inequality.
\end{abstract}
}


\section{Introduction}

Interactive statistical decision making (ISDM) \cite{chen2024assouad} studies learning problems in which the learner's actions affect what data are observed.
A model \(M\in\mathcal M\) specifies an environment, the learner chooses an algorithm \(\textup{ALG}\in\mathcal D\), and the resulting data \(X\) are generated according to an \emph{algorithm-dependent} law
\(
X \sim P^{M,\textup{ALG}}.
\)
This coupling between data and the algorithm sharply contrasts with classical (non-interactive) estimation, where one observes i.i.d.\ samples from a distribution \(P_\theta\) that does not depend on the estimator.
ISDM encompasses a broad range of problems, including multi-armed bandits, contextual bandits, and reinforcement learning (RL) \cite{lattimore2020bandit,sutton2018reinforcement}.
For example, in a stochastic \(K\)-armed bandit, pulling arm \(k\) reveals only a reward from arm \(k\)'s distribution, and reveals nothing directly about the other arms.
As a result, the learner must balance \emph{exploration} (gathering information) and \emph{exploitation} (using the best-known action), which drives both algorithm design and information-theoretic limits \cite{lai1985asymptotically,auer2002finite}.

Lower bounds formalize fundamental limits of learning in ISDM.
In a Bayesian formulation, one posits a prior \(\mu\) over instances \(M\), and evaluates performance under the induced prior-predictive law \((M,X)\sim \mu(dM)P^{M,\textup{ALG}}(dX)\).
A Bayesian lower bound asserts that, for any algorithm \(\textup{ALG}\), the induced risk cannot be below a certain threshold.
Proving such results is subtle in ISDM because the information available to the learner depends on \(\textup{ALG}\) itself.

In non-interactive estimation, classical techniques such as Fano's method, Le Cam's method, and Assouad's lemma provide general-purpose tools to lower bound Bayesian and minimax risks \cite{tsybakov2009introduction}.
In interactive settings, sharp lower bounds were historically developed case-by-case (e.g., for bandits and RL) \cite{lai1985asymptotically,auer2002finite,jaksch2010near,jin2018q,osband2016lower}.
Foster et al.\ introduced the \emph{Decision--Estimation Coefficient} (DEC), which yields general minimax lower bounds for regret and sample complexity
\cite{foster2021statistical,foster2023tight}.
Chen et al.\ proposed the \emph{interactive Fano method}, providing a unifying view that connects classical tools (Fano/Le Cam/Assouad) with the DEC approach and introducing fractional covering numbers to tighten bounds for convex model classes \cite{chen2024assouad}.

Most of the above literature focuses on lower bounds for \emph{expected} risk (or expected regret).
However, expectation can obscure rare but consequential failures, which are central in safety- and reliability-motivated applications.
One way to capture tails is through \emph{quantile} risk.
Chen et al.\ derive interactive lower bounds for quantile-type criteria within their framework \cite{chen2024assouad}, and Ma et al.\ develop risk-level-explicit minimax quantile lower bounds for non-interactive problems \cite{ma2024highprobability}.
(Their minimax quantile criterion is equivalent to minimax VaR of the loss, via \(\mathbb{P}(L>r)\le \delta \iff \mathbb{P}(L\le r)\ge 1-\delta\). It coincides with the standard definitions of VaR in risk management \cite{rockafellar2000optimization,rockafellar2002general}.)
Moreover, CVaR (also known as Expected Shortfall) satisfies \(\textup{CVaR}_\alpha(L)\ge \textup{VaR}_\alpha(L)\) for losses, so VaR lower bounds imply CVaR lower bounds at the same level \cite{rockafellar2000optimization,rockafellar2002general,acerbi2002expected}.

In this work, we depart from the expected-risk setting and generalize the interactive Fano approach of \cite{chen2024assouad}.
Rather than specializing to a single quantile event, we develop a lower-bound template that applies to a broad class of bounded loss functionals.
Our framework extends the Fano inequality in interactive settings, and it also recovers non-interactive lower bounds as a special case.
As a consequence, we obtain lower bounds for \emph{Bayesian} CVaR of the loss under the prior-predictive mixture induced by \(\mu\) and \(\textup{ALG}\).
This perspective differs from minimax quantile/VaR analyses such as \cite{ma2024highprobability} in two key ways: (i) we directly target the prior-predictive (Bayesian) tail risk rather than the worst-case (minimax) tail risk, and (ii) our bounds apply to interactive decision making problems.

CVaR is a widely used tail-risk measure in finance and risk management \cite{rockafellar2000optimization,rockafellar2002general,acerbi2002expected}.
It has also appeared in sequential decision making, for instance in risk-sensitive RL and bandits, often through algorithmic objectives and policy-gradient style methods \cite{chow2014cvar,tamar2015coherent}.
Our focus here is different: we study \emph{information-theoretic lower bounds} for Bayesian CVaR in interactive learning, yielding fundamental limits that complement algorithmic developments.

\section{Our Contributions}
\begin{itemize}
    \item We generalize the interactive Fano method of \cite{chen2024assouad} from a single success/quantile event to \emph{arbitrary bounded transforms} of the loss via a randomized one-bit statistic.

    \item We invert this Bernoulli \(f\)-divergence to obtain a \emph{two-sided} bound for the expected transform. This recovers the one-sided tail/quantile bound of \cite{chen2024assouad} as a special case and provides a direct way to lower bound different loss functionals across any ISDM problem.

    \item Using the hinge transform together with the Rockafellar--Uryasev representation \cite{rockafellar2000optimization,rockafellar2002general}, we derive lower bounds on the prior-predictive (Bayesian) \(\textup{CVaR}_\alpha\) of bounded losses, applicable across ISDM and including standard estimation as a special case.

    \item For the KL divergence and the mixture reference distribution, we obtain an explicit Bayesian \(\textup{CVaR}\) lower bound in terms of mutual information (via Pinsker), yielding a simple closed-form converse for interactive problems.
\end{itemize}

\section{Notation}
Random variables are denoted by capitals (e.g., \(M,X\)); realizations are lower case. We write ``a.s.'' for almost surely.
For a random variable \(Z\), we write \(\textup{Law}(Z)\) for its distribution, and \(\textup{Law}_P(Z)\) when we wish to emphasize that the underlying law of the relevant inputs is \(P\).

We work with measurable spaces \((\mathcal X,\mathcal F)\), \((\mathcal U,\mathcal H)\), and \((\mathcal Y,\mathcal G)\).
A Markov kernel \(K\) from \((\mathcal X,\mathcal F)\) to \((\mathcal Y,\mathcal G)\) is a map
\(K:\mathcal X\times\mathcal G\to[0,1]\) such that for each \(x\in\mathcal X\), \(A\mapsto K(A\mid x)\) is a probability measure on \((\mathcal Y,\mathcal G)\), and for each \(A\in\mathcal G\), \(x\mapsto K(A\mid x)\) is \(\mathcal F\)-measurable.
For a probability measure \(P\) on \((\mathcal X,\mathcal F)\), the output measure \(K\circ P\) on \((\mathcal Y,\mathcal G)\) is
\[
(K\circ P)(A)\;:=\;\int_{\mathcal X} K(A\mid x)\,P(dx),\qquad A\in\mathcal G.
\]

We represent randomization by an auxiliary random variable \(U\sim\nu\) on \((\mathcal U,\mathcal H)\), assumed independent of all other random variables under consideration unless stated otherwise.
Given a measurable map \(\Phi:(\mathcal X\times\mathcal U,\mathcal F\otimes\mathcal H)\to(\mathcal Y,\mathcal G)\), define the induced Markov kernel \(K\) by
\[
K(A\mid x)\;:=\;\int_{\mathcal U}\mathbf 1\{\Phi(x,u)\in A\}\,\nu(du),
\qquad x\in\mathcal X,\; A\in\mathcal G.
\]
If \((X,U)\sim P\otimes \nu\), then \(K\circ P=\textup{Law}_{P\otimes \nu}\big(\Phi(X,U)\big)\). We write
\(
K\circ P \;=\; \textup{Law}_P\big(\Phi(X,U)\big),
\) indicating that \(X\sim P\) and \(U\sim\nu\) is independent, and we want to emphasize the dependence of $\Phi(X,U)$ on $P$.

For each model \(m\in\mathcal M\) and algorithm \(\textup{ALG}\), let \({P}^{m,\textup{ALG}}\) denote the induced law of the observation \(X\) on \((\mathcal X,\mathcal F)\); for measurable \(A\in\mathcal F\), we write
\(
{P}^{m,\textup{ALG}}(A)\;:=\;\mathbb{P}(X\in A \mid M=m,\textup{ALG}),
\) and
\(
\mathbb{E}^{m,\textup{ALG}}[h(X)]\;:=\;\mathbb{E}[h(X)\mid M=m,\textup{ALG}],
\)
for any measurable \(h\) with finite expectation.
We view \(m\mapsto {P}^{m,\textup{ALG}}\) as a Markov kernel from \((\mathcal M,\Sigma_{\mathcal M})\) to \((\mathcal X,\mathcal F)\), and we write
\({P}^{M,\textup{ALG}}(\cdot):=\mathbb{P}(X\in\cdot\mid M,\textup{ALG})\) a.s.

Given a prior \(\mu\in\Delta(\mathcal M)\), we define the mixture law and expectation by
\(
\mathbb{P}_{\mu,\textup{ALG}}(\cdot)\;:=\;\mathbb{E}_{M\sim\mu}\big[{P}^{M,\textup{ALG}}(\cdot)\big]
\) and
\(
\mathbb{E}_{\mu,\textup{ALG}}[\cdot]\;:=\;\mathbb{E}_{M\sim\mu}\big[\mathbb{E}^{M,\textup{ALG}}[\cdot]\big].
\)
Indicators are denoted \(\mathbf 1\{\cdot\}\), and \([z]_+:=\max\{z,0\}\).
Throughout, we may take \(U\sim\mathrm{Unif}[0,1]\) as a default auxiliary random variable (independent of all other random variables under consideration) unless noted otherwise.
For probability measures \(P,Q\) on a common measurable space and a convex function \(f:\mathbb R_+\to\mathbb R\) with \(f(1)=0\), the \(f\)-divergence is
\(
D_f(P\|Q)\;:=\;\int q\, f\!\left(\frac{p}{q}\right)\,d\lambda,
\)
where \(\lambda\) is any dominating measure and \(p=\frac{dP}{d\lambda}\), \(q=\frac{dQ}{d\lambda}\) are Radon--Nikodym derivatives.
We write \(\mathrm{Bern}(p)\) for the Bernoulli distribution with mean \(p\).
For $f=\mathrm{KL}$ we write, \(D_{\mathrm{KL}}(P\|Q)=\int p\log\!\big(p/q\big)\,d\lambda\), and with the mixture \(Q^\star=\mathbb{E}_{M\sim\mu}P^{M,\textup{ALG}}\) we define the mutual information \(I_{\mu,\textup{ALG}}(M;X):=\mathbb{E}_{M\sim\mu}D_{\mathrm{KL}}(P^{M,\textup{ALG}}\|Q^\star)\).

\section{Problem setup}
We adopt the formulation of \cite{chen2024assouad}, which treats classical statistical estimation and interactive decision problems within a single framework. Let $\mathcal X$ denote the space of all possible observations (including full interaction transcripts), let $\mathcal M$ be a class of environments/models, and let $\mathcal D$ be a class of algorithms (decision rules, policies, or procedures). For each pair $(M,\textup{ALG})\in\mathcal M\times\mathcal D$, the interaction between $\textup{ALG}$ and the environment $M$ induces a probability law $\mathbb P^{M,\textup{ALG}}$ over $X\in\mathcal X$, and we write
\(
X \sim  P^{M,\textup{ALG}}.
\)
A nonnegative loss functional $L:\mathcal M\times\mathcal X\to\mathbb R_+$ evaluates the performance of $\textup{ALG}$ in model $M$ on the outcome $X$. An ISDM instance is thus specified by the quadruple $(\mathcal X,\mathcal M,\mathcal D,L)$, and the expected risk of $\textup{ALG}$ under $M$ is
\(
\mathbb E^{M,\textup{ALG}}\!\bigl[L(M,X)\bigr].
\)
Typically one measures the performance in the worst case over models via the minimax risk
\(
\mathfrak M
~:=~
\inf_{\textup{ALG}\in\mathcal D}\;
\sup_{M\in\mathcal M}\;
\mathbb E^{M,\textup{ALG}}\!\bigl[L(M,X)\bigr]
\)
or the Bayes risk given a prior 
\(\mu\in\Delta(\mathcal M)\): \(\mathfrak{F}(\mu)~:=~
\inf_{\textup{ALG}\in\mathcal D}\mathbb{E}_{M\sim\mu}\mathbb E^{M,\textup{ALG}}\!\bigl[L(M,X)]\). 

Chen et al. \cite{chen2024assouad} derive an interactive Fano method providing a lower bound for the quantile \(\mathbb{P}_{M\sim\mu,\;X\sim P^{M,\mathrm{ALG}}}\!\bigl[L(M,X)]\). Using this quantile lower bound, they subsequently derive a lower bound for the Bayes risk and the minimax risk.

In this work, we generalize the Fano method further, deriving lower bounds for any expected bounded transform of the loss $\mathbb{E}[\phi(L)]$ (generalizing quantile lower bounds) and provide lower bounds for $\textup{CVaR}(L)$) under $\mu\otimes P^{M,\textup{ALG}}$ to move beyond expected risk criteria.

\section{Prior results}
Chen et al.~\cite{chen2024assouad} established an interactive Fano inequality by combining data processing for \(f\)-divergences with a Bernoulli monotonicity argument. In this section, we re-derive their result with an eye toward generalization, isolating the proof steps that can be extended and those that are specific to the original formulation.

\begin{lemma}[Data processing inequality for \(f\)-divergences \cite{polyanskiy2025information}]\label{lem:dpi-f}
Let \((\mathcal X,\mathcal F)\), \((\mathcal U,\mathcal H)\), and \((\mathcal Y,\mathcal G)\) be measurable spaces.
Let \(P_X,Q_X\) be probability measures on \((\mathcal X,\mathcal F)\), and let \(\nu\) be a probability measure on \((\mathcal U,\mathcal H)\).
Let \(\Phi:(\mathcal X\times\mathcal U,\mathcal F\otimes\mathcal H)\to(\mathcal Y,\mathcal G)\) be measurable.

Define a Markov kernel \(T:\mathcal X\times\mathcal G\to[0,1]\) by
\[
T(A\mid x)\;:=\;\int_{\mathcal U}\mathbf 1\{\Phi(x,u)\in A\}\,\nu(du),
\qquad x\in\mathcal X,\; A\in\mathcal G.
\]
Let \(T\circ P_X\) denote the pushforward (output) measure on \((\mathcal Y,\mathcal G)\) induced by \(P_X\) through \(T\),
\[
(T\circ P_X)(A)\;:=\;\int_{\mathcal X}T(A\mid x)\,P_X(dx),\qquad A\in\mathcal G,
\]
and similarly define \(T\circ Q_X\).

Then for any \(f\)-divergence \(D_f\),
\[
D_f(P_X\|Q_X)\ \ge\ D_f\big(T\circ P_X\ \|\ T\circ Q_X\big).
\]
Equivalently, if \((X,U)\sim P_X\otimes \nu\) and \((X',U)\sim Q_X\otimes \nu\), then
\[
T\circ P_X \;=\; \textup{Law}_{P_X\otimes \nu}\big(\Phi(X,U)\big)
\]
\[
T\circ Q_X \;=\; \textup{Law}_{Q_X\otimes \nu}\big(\Phi(X',U)\big),
\]
and hence
\[
D_f(P_X\|Q_X)\ \ge\ \]
\[
D_f\!\Big(\textup{Law}_{P_X\otimes \nu}\big(\Phi(X,U)\big)\ \Big\|\ 
\textup{Law}_{Q_X\otimes \nu}\big(\Phi(X',U)\big)\Big).
\]
\end{lemma}

Chen et al.~\cite{chen2024assouad} prove the following lemma.

\begin{lemma}[Bernoulli monotonicity]
\label{lem:bern-mono}
For fixed \(y\in[0,1]\), the map \(x\mapsto D_f(\mathrm{Bern}(x)\,\|\,\mathrm{Bern}(y))\)
is nondecreasing on \([y,1]\) and nonincreasing on \([0,y]\).
\end{lemma}
Based on the above lemmas, Chen et al.~\cite{chen2024assouad} prove an interactive Fano inequality
which says if, for some reference distribution $Q$, the average $f$-divergence
\(
\mathbb{E}_{M\sim\mu}\!\left[D_f\!\left(P^{M,\mathrm{ALG}}\,\|\,Q\right)\right]
\)
stays below a certain threshold $d_{f,\delta}\!\left(\rho_{\Delta,Q}\right)$, then the algorithm must incur loss at least $\Delta$ with probability at least $\delta$, i.e.,
\[
\mathbb{E}_{M\sim\mu}\!\left[D_f\!\left(P^{M,\mathrm{ALG}}\,\|\,Q\right)\right]
< d_{f,\delta}\!\left(\rho_{\Delta,Q}\right)
\]
\[
\quad\Longrightarrow\quad
\mathbb{P}_{M\sim\mu,\;X\sim P^{M,\mathrm{ALG}}}\!\big(L(M,X)\ge \Delta\big)\ge \delta.
\]
\begin{theorem}[Interactive quantile Fano]
\label{thm:Fano-expected-risk}
Fix an \(f\)-divergence \(D_f\). Let \(\textup{ALG}\) be a given algorithm, let \(\mu\in\Delta(\mathcal M)\) be a prior, and let \(\Delta>0\) be a risk level. For any reference \(Q\in\Delta(\mathcal X)\) define
\(
\rho_{\Delta,Q}\;:=\; \mathbb P_{M\sim\mu,\; X\sim Q}\!\big(L(M,X)<\Delta\big).
\)
Then we have the following quantile lower bound 
\[
\begin{aligned}
&\mathbb P_{M\sim\mu,\; X\sim P^{M,\textup{ALG}}}\!\big(L(M,X)\ge \Delta\big)
\ge
\\
&\sup_{Q\in\Delta(\mathcal X),\; \delta\in[0,1]} \{\,
\delta \;:\; \mathbb{E}_{M\sim\mu}\!\big[D_f\!\big(P^{M,\textup{ALG}}\,\|\,Q\big)\big]
\!<\! d_{f,\delta}\!\big(\rho_{\Delta,Q}\big)
\,\},
\end{aligned}
\]
where
\[
d_{f,\delta}(p)\;:=\;
\begin{cases}
D_f\!\big(\mathrm{Bern}(1-\delta)\,\big\|\,\mathrm{Bern}(p)\big), & \text{if } p\le 1-\delta,\\[4pt]
0, & \text{otherwise.}
\end{cases}
\]
\end{theorem}

\begin{proof}[Proof]
We work with the two joint laws
\[
P_0(m,x)\;=\;\mu(m)P^{M,\textup{ALG}}(x), \qquad 
P_1(m,x)\;=\;\mu(m)Q(x).
\]
The argument proceeds in three steps.

\emph{i) (Indicator reduction).}
Compress \((M,X)\) to the success indicator
\[
S\;:=\;\mathbf 1\{L(M,X)<\Delta\}.
\]
Then, for \(i\in\{0,1\}\),
\[
\mathbb P_{(M,X)\sim P_i}(S=1)
= \mathbb P_{(M,X)\sim P_i}\!\big(L(M,X)<\Delta\big).
\]
Hence, under \(P_0\),
\[
S\ \sim\ \mathrm{Bern}(\bar\rho_\Delta),
\qquad 
\bar\rho_\Delta\;:=\;P_0\!\big(L(M,X)<\Delta\big),
\]
and under \(P_1\),
\[
S\ \sim\ \mathrm{Bern}(\rho_{\Delta,Q}),
\qquad 
\rho_{\Delta,Q}\;:=\;P_1\!\big(L(M,X)<\Delta\big).
\]
Equivalently,
\[
\mathsf{Law}_{P_0}(S)=\mathrm{Bern}(\bar\rho_\Delta),
\qquad
\mathsf{Law}_{P_1}(S)=\mathrm{Bern}(\rho_{\Delta,Q}).
\]

\emph{ii) (Data processing).}
By the data-processing inequality for \(f\)-divergences applied to the measurable map \((m,x)\mapsto S\),
\begin{equation}
\label{eq:dp-bern}
\begin{aligned}
    &D_f\!\big(\mathrm{Bern}(\bar\rho_\Delta)\,\big\|\,\mathrm{Bern}(\rho_{\Delta,Q})\big)
    \ \le\
    D_f(P_0\|P_1)
    \\
    &=\
    \mathbb{E}_{M\sim\mu}\, D_f\!\big(P^{M,\textup{ALG}}\,\|\,Q\big).
\end{aligned}
\end{equation}

\emph{iii) (One-sided Inversion).}

Fix \(Q\) and \(\delta\in[0,1]\) such that
\[
\begin{aligned}
&\mathbb{E}_{M\sim\mu}\, D_f\!\big(P^{M,\textup{ALG}}\,\|\,Q\big)
\ <\ d_{f,\delta}(\rho_{\Delta,Q})
\\
&=
\begin{cases}
D_f\!\big(\mathrm{Bern}(1-\delta)\,\big\|\,\mathrm{Bern}(\rho_{\Delta,Q})\big), & \rho_{\Delta,Q}\le 1-\delta,\\
0, & \rho_{\Delta,Q}>1-\delta.
\end{cases}
\end{aligned}
\]
Since the left-hand side is nonnegative, the strict inequality forces \(\rho_{\Delta,Q}<1-\delta\).
Combining with \eqref{eq:dp-bern} yields
\[
D_f\!\big(\mathrm{Bern}(\bar\rho_\Delta)\big\|\,\mathrm{Bern}(\rho_{\Delta,Q})\big)
\!\!<\!\!\!\
D_f\!\big(\mathrm{Bern}(1-\delta)\,\big\|\,\mathrm{Bern}(\rho_{\Delta,Q})\big).
\]

\emph{Case 1:} \(\bar\rho_\Delta \le \rho_{\Delta,Q}\).
Then \(\bar\rho_\Delta \le \rho_{\Delta,Q} < 1-\delta\), hence directly \(\bar\rho_\Delta<1-\delta\).

\emph{Case 2:} \(\bar\rho_\Delta > \rho_{\Delta,Q}\).
Using Lemma~\ref{lem:bern-mono} with \(y=\rho_{\Delta,Q}\) and \(1-\delta\ge \rho_{\Delta,Q}\), the strict inequality above implies
\(\bar\rho_\Delta < 1-\delta\).

In both cases we conclude \(\bar\rho_\Delta<1-\delta\), i.e.,
\[
\mathbb P_{M\sim\mu,\,X\sim P^{M,\textup{ALG}}}\!\big(L(M,X)\ge \Delta\big)
= 1-\bar\rho_\Delta > \delta.
\]
Taking the supremum over all such pairs \((Q,\delta)\) completes the proof.
\end{proof}
\begin{remark}[Discussion: proof steps in the interactive quantile Fano]
The interactive Fano argument for quantile risk can be organized into three conceptual steps:
(i) an \emph{indicator reduction} that compresses the loss into an indicator,
(ii) a \emph{data-processing} bound,
and (iii) an \emph{inversion} step that translates the bound into a lower bound on the risk.
This decomposition also indicates how to extend the method. For more general loss functionals, one must replace the indicator reduction and the resulting \emph{one-sided} inversion by a construction that supports a \emph{two-sided} inversion.
\end{remark}

The tail risk bound gives rise to an in-expectation bound from Markov's inequality.
\begin{remark}[From tail to expectation]
For any nonnegative loss, \(\mathbb{E}[L]\ge \Delta\,\mathbb P(L\ge \Delta)\), so Theorem~\ref{thm:Fano-expected-risk}
immediately implies an \emph{expected-risk} lower bound:
\[
\begin{aligned}
&\mathbb{E}_{M\sim\mu,\,X\sim P^{M,\textup{ALG}}}[L(M,X)]
\ \ge\
\\
&\Delta\cdot
\sup_{Q,\delta}\Big\{\delta:\ \mathbb{E}_M D_f(P^{M,\textup{ALG}}\|Q)< d_{f,\delta}(\rho_{\Delta,Q})\Big\}.
\end{aligned}
\]
\end{remark}

\section{New results}
Motivated by \cite{chen2024assouad}, we extend Theorem~\ref{thm:Fano-expected-risk} in two ways. 
First, we replace the single quantile event with a general criterion of the form $\mathbb{E}[\phi(L)]$ for any bounded function $\phi$. 
Second, we refine the Bernoulli inversion step to obtain a two-sided bound, which is essential for controlling more general loss functionals.

\begin{theorem}[Generalized Interactive Fano inequality]
\label{thm:bern-ball}
Fix $D_f,\mu,\textup{ALG},L$ as above, any $Q\in\Delta(\mathcal X)$, and any bounded function $\phi:\mathbb R\to[0,1]$ (more generally, any bounded $\phi$ after affine rescaling to $[0,1]$).
 Let
\[
\bar\rho_\phi
:=
\mathbb E_{M\sim\mu,\,X\sim P^{M,\textup{ALG}}}\big[\phi(L(M,X))\big]\]
\[
\rho_{\phi,Q}
:=
\mathbb E_{M\sim\mu,\,X\sim Q}\big[\phi(L(M,X))\big],
\]
where in both expectations $M$ and $X$ are independent given their laws, and set
\(
B
:=
\mathbb E_{M\sim\mu}\, D_f\big(P^{M,\textup{ALG}}\big\|Q\big).
\)
Then
\(
D_f\!\big(\mathrm{Bern}(\bar\rho_\phi)\,\big\|\,\mathrm{Bern}(\rho_{\phi,Q})\big)\ \le\ B.
\)
Consequently, with
\[
a_f^{-}(B;b):=\inf\{a\in[0,1]:\ D_f(\mathrm{Bern}(a)\|\mathrm{Bern}(b))\le B\}
\]
and
\[
a_f^{+}(B;b):=\sup\{a\in[0,1]:\ D_f(\mathrm{Bern}(a)\|\mathrm{Bern}(b))\le B\},
\]
we have the two-sided bound
\[
a_f^{-}(B;\rho_{\phi,Q})\ \le\ \bar\rho_\phi\ \le\ a_f^{+}(B;\rho_{\phi,Q}).
\]
\end{theorem}

\begin{proof}
Define the two joint laws on $(\mathcal M\times\mathcal X)$ by
\[
P_0(m,x):=\mu(m)\,P^{M,\textup{ALG}}(x),
\qquad
P_1(m,x):=\mu(m)\,Q(x).
\]

\emph{(Indicator reduction).}
Let $U\sim\mathrm{Unif}[0,1]$ be independent of $(M,X)$ under both $P_0$ and $P_1$, and set the randomized one–bit statistic
\(
Y:=\mathbf 1\{U\le \phi(L(M,X))\}.
\)
Write $Z:=\phi(L(M,X))\in[0,1]$. Then, for $i\in\{0,1\}$,
\[
\begin{aligned}
\mathbb P_{P_i}(Y=1)
&=\mathbb P_{P_i}\big(U\le \phi(L(M,X))\big)
\\
&=\mathbb E_{P_i}\big[\mathbf 1\{U\le \phi(L(M,X))\}\big]
\\
&=\mathbb E_{P_i}\big[\mathbb P(U\le Z\mid Z)\big]
\;=\;\mathbb E_{P_i}[Z],
\end{aligned}
\]
where we used that $U$ is independent of $(M,X)$ and uniformly distributed on $[0,1]$, so that
$\mathbb P(U\le z\mid Z=z)=z$ for $z\in[0,1]$.
By the definitions of $\bar\rho_\phi$ and $\rho_{\phi,Q}$, this implies
\[
\mathbb P_{P_0}(Y=1)=\bar\rho_\phi,
\qquad
\mathbb P_{P_1}(Y=1)=\rho_{\phi,Q},
\]
so that
\(
Y \sim \mathrm{Bern}(\bar\rho_\phi)\ \text{under }P_0,
\qquad
Y \sim \mathrm{Bern}(\rho_{\phi,Q})\ \text{under }P_1.
\)

\emph{(Data processing).}
By the data–processing inequality for $f$–divergences applied to the measurable function $(m,x)\rightarrow Y$
we obtain,
\[
\begin{aligned}
D_f(P_0\|P_1)
&\ \ge\ D_f\!\big(\mathsf{Law}_{P_0}(Y)\,\big\|\,\mathsf{Law}_{P_1}(Y)\big)
\\
&\ =\ D_f\!\big(\mathrm{Bern}(\bar\rho_\phi)\,\big\|\,\mathrm{Bern}(\rho_{\phi,Q})\big).
\end{aligned}
\]
Moreover, using the product form of $P_0$ and $P_1$ and the definition of the $f$–divergence, we have
\[
\begin{aligned}
D_f(P_0\|P_1)
&=
\int_{\mathcal M}
D_f\!\big(P^{M,\textup{ALG}}\big\|Q\big)\,\mu(dm)
\\
&=
\mathbb E_{M\sim\mu}\, D_f\!\big(P^{M,\textup{ALG}}\big\|Q\big)
=: B.
\end{aligned}
\]
Combining these two yields
\[
D_f\!\big(\mathrm{Bern}(\bar\rho_\phi)\,\big\|\,\mathrm{Bern}(\rho_{\phi,Q})\big)\ \le\ B.
\]

\emph{(Two–sided bound).}
For fixed $b\in[0,1]$, the map $a\mapsto D_f(\mathrm{Bern}(a)\|\mathrm{Bern}(b))$ is continuous and convex on $[0,1]$; hence the sublevel set
\(
\{a\in[0,1]:\ D_f(\mathrm{Bern}(a)\|\mathrm{Bern}(b))\le B\}
\)
is a closed interval $[a_f^{-}(B;b),\,a_f^{+}(B;b)]$, since the preimage of a closed set under a continuous function is closed. Since $\bar\rho_\phi$ satisfies the inequality with $b=\rho_{\phi,Q}$, it follows that 
\(
a_f^{-}(B;\rho_{\phi,Q})\ \le\ \bar\rho_\phi\ \le\ a_f^{+}(B;\rho_{\phi,Q}).
\)\end{proof}

Theorem \ref{thm:bern-ball} says that if the average divergence
\(B=\mathbb{E}_{M\sim\mu} D_f(P^{M,\textup{ALG}}\|Q)\) is small i.e., the real
transcript is nearly indistinguishable from the benchmark \(Q\) then, the transformed risk \(\bar\rho_\phi=\mathbb{E}[\phi(L)]\) cannot
deviate much from its benchmark value \(\rho_{\phi,Q}\). The theorem quantifies this closeness via the two-sided bound.
In particular, with the mixture benchmark \(Q^\star=\mathbb{E}_{M\sim\mu} P^{M,\textup{ALG}}\) and \(f=\mathrm{KL}\), we have \(B=I_{\mu,\textup{ALG}}(M;X)\).
The theorem says that if we have low mutual information,  where \(X\) carries little information about \(M\), then the (transformed) risk \(\bar\rho_\phi\) remains close to the benchmark \(\rho_{\phi,Q^\star}\).
\begin{remark}
Chen et al.\ use the indicator \(\mathbf 1\{L(M,X)<\Delta\}\), whereas we generalize to the randomized bit
\(\mathbf 1\{U\le \phi(L(M,X))\}\).
If we specialize our indicator to \(\phi(l)=\mathbf 1\{l<\Delta\}\) as in \cite{chen2024assouad} we recover the indicator.
\[
Y=\mathbf 1\{U\le \phi(L(M,X))\}
=\mathbf 1\{U\le \mathbf 1(L(M,X)<\Delta)\}.
\]
Since \(\mathbb P(U=0)=0\), it follows that
\[
Y=
\begin{cases}
1, & L(M,X)<\Delta,\\
0, & L(M,X)\ge \Delta,
\end{cases}
\qquad \text{a.s.}
\]
Hence, \(Y=\mathbf 1\{L(M,X)<\Delta\}\) a.s.
\end{remark}

\begin{remark}[Two-sided vs Chen et al.'s bound]
Chen et al.\ \cite{chen2024assouad} compare
\(
B:=\mathbb E_{M\sim\mu}\, D_f\!\big(P^{M,\textup{ALG}}\big\|Q\big)
\)
to the Bernoulli calibration term \(d_{f,\delta}\!\big(\rho_{\Delta,Q}\big)\) and, via a two–case monotonicity argument, deduce an \emph{upper} bound on
\(\mathbb P_{M\sim\mu,\,X\sim P^{M,\textup{ALG}}}\!\big(L(M,X)<\Delta\big)\), equivalently a \emph{lower} bound on the tail \(\mathbb P(L\ge\Delta)\).
In contrast, we \emph{invert} the Bernoulli \(f\)-ball
\(
D_f\!\big(\mathrm{Bern}(\bar\rho_\phi)\,\big\|\,\mathrm{Bern}(\rho_{\phi,Q})\big)\ \le\ B
\)
to obtain a two–sided interval
\(\bar\rho_\phi\in\big[a_f^{-}(B;\rho_{\phi,Q}),\,a_f^{+}(B;\rho_{\phi,Q})\big]\),
where \(\bar\rho_\phi=\mathbb E[\phi(L(M,X))]\).
This yields converse bounds for a variery of functionals:
(i) for \emph{nonincreasing} transforms (e.g., \(\phi(l)=\mathbf 1\{l<\Delta\}\)),
an \emph{upper} bound on \(\bar\rho_\phi\) implies a \emph{lower} bound on the quantile/tail risk,
\(
\mathbb P(L\ge\Delta)=1-\mathbb E[\phi(L)];
\)
(ii) for \emph{nondecreasing} transforms (e.g., the hinge \(\phi_t(l)=(l-t)_+/L_{\max}\) used for CVaR),
the \emph{lower} endpoint \(a_f^{-}\) delivers \emph{lower} bounds on \(\mathbb E[\phi(L)]\), which translate to CVaR lower bounds via the Rockafellar–Uryasev identity.
Many risks of interest (mean, CVaR, quantiles via tails) are monotone nondecreasing in \(L\) and we can use the two–sided inversion therefore to target the correct direction for converse bounds.
\end{remark}
Using Theorem \ref{thm:bern-ball}, we derive the following one-sided bound.
\begin{corollary}[One-sided bounded-transform from the Generalized Fano method]
\label{cor:bt-interactive-Fano-calibrated}
Assume the setting of Theorem~\ref{thm:bern-ball}. For any measurable
$\phi:[0,\infty)\to[0,1]$ and any $Q\in\Delta(\mathcal X)$ define
\(
\bar\rho_\phi
:= \mathbb{E}_{M\sim\mu,\,X\sim P^{M,\textup{ALG}}}\!\big[\phi(L(M,X))\big],
\) 
\(
\rho_{\phi,Q}
:= \mathbb{E}_{M\sim\mu,\,X\sim Q}\!\big[\phi(L(M,X))\big],
\)
and for $\theta,p\in[0,1]$ set
\[
d_{f,\theta}(p)
:=
\begin{cases}
D_f\!\big(\mathrm{Bern}(\theta)\,\big\|\,\mathrm{Bern}(p)\big), & \theta\ge p,\\
0, & \theta<p.
\end{cases}
\]
Then \(\bar\rho_\phi
\;\le\;\)
\[
\inf_{Q\in\Delta(\mathcal X),\, \theta\in[0,1]}\;
\Big\{\theta:
\ \mathbb{E}_{M\sim\mu} D_f\!\big(P^{M,\textup{ALG}}\,\big\|\,Q\big)
\ <\ d_{f,\theta}\!\big(\rho_{\phi,Q}\big)\Big\}.
\]
\end{corollary}

\begin{proof}
Fix $Q\in\Delta(\mathcal X)$ and $\theta\in[0,1]$ such that
\(
\mathbb{E}_{M\sim\mu} D_f\!\big(P^{M,\textup{ALG}}\,\big\|\,Q\big)\ <\ d_{f,\theta}\!\big(\rho_{\phi,Q}\big).
\)
By Theorem~\ref{thm:bern-ball},
\(
D_f\!\big(\mathrm{Bern}(\bar\rho_\phi)\,\big\|\,\mathrm{Bern}(\rho_{\phi,Q})\big)\)
\( \le\
\mathbb{E}_{M\sim\mu} D_f\!\big(P^{M,\textup{ALG}}\,\big\|\,Q\big)
<\
d_{f,\theta}\!\big(\rho_{\phi,Q}\big).
\)
Since $d_{f,\theta}(\rho_{\phi,Q})=0$ when $\theta<\rho_{\phi,Q}$, the strict inequality forces
$\theta\ge\rho_{\phi,Q}$, and thus
\(
D_f\!\big(\mathrm{Bern}(\bar\rho_\phi)\,\big\|\,\mathrm{Bern}(\rho_{\phi,Q})\big)
\ <\
D_f\!\big(\mathrm{Bern}(\theta)\,\big\|\,\mathrm{Bern}(\rho_{\phi,Q})\big).
\)
For fixed $y\in[0,1]$, the map $x\mapsto D_f(\mathrm{Bern}(x)\,\|\,\mathrm{Bern}(y))$
is nondecreasing on $[y,1]$ and nonincreasing on $[0,y]$; hence the strict inequality implies
$\bar\rho_\phi\le \theta$ (if $\bar\rho_\phi\le y$ then $\bar\rho_\phi\le y\le \theta$, and if
$\bar\rho_\phi> y$ monotonicity on $[y,1]$ gives $\bar\rho_\phi<\theta$).
Taking the infimum over all admissible pairs $(Q,\theta)$ yields the claim.
\end{proof}
\begin{remark}[Recovery of Chen--et al.\ Thm.~2]
Taking $\phi(l)=\mathbf 1\{l<\Delta\}$ gives
$\bar\rho_\phi=\mathbb P(L<\Delta)$ and using Corollary \ref{cor:bt-interactive-Fano-calibrated} with $\theta = 1-\delta$ reduces to the interactive Fano inequality of \cite{chen2024assouad},
yielding a lower bound on the tail probability $\mathbb P(L\ge \Delta)$.
\end{remark}

Using suitable choices of \(\phi\), Theorem \ref{thm:bern-ball} can yield converses for other risks e.g., CVaR, truncated means via \(\phi(l)=\min\{l/\tau,1\}\) giving bounds for $(1/\tau)\mathbb{E}[\min\{l,\tau\}]$, and the laplace transform risk $\mathbb{E}[e^{-\lambda l}]$ via \(\phi(l)=e^{-\lambda l}\) for $\tau, \lambda > 0$. We  prove a lower bound for the CVaR building block below.

\begin{corollary}[Building block for CVaR lower bounds (from the generalized Fano)]
\label{cor:hinge-lower}
Assume $0\le L(M,X)\le L_{\max}$ and fix $t\in\mathbb R$. Define the bounded hinge
\(
\phi_t(l):=\frac{(l-t)_+}{L_{\max}}\in[0,1] \text{ with }
a_t:=\mathbb{E}_{\mu,\,P^{M,\textup{ALG}}}\!\big[\phi_t(L)\big]
\) and \(
b_t:=\mathbb{E}_{\mu,\,Q}\!\big[\phi_t(L)\big]=\frac{1}{L_{\max}}\mathbb{E}_{\mu,\,Q}[(L-t)_+].
\)
Let $B:=\mathbb{E}_{M\sim\mu}D_f(P^{M,\textup{ALG}}\|Q)$ and
\(
a_f^{-}(B;b)\ :=\ \inf\big\{a\in[0,1]:\ D_f\big(\mathrm{Bern}(a)\,\big\|\,\mathrm{Bern}(b)\big)\le B\big\}.
\)
Then Theorem~\ref{thm:bern-ball} yields
\(
\mathbb{E}_{\mu,\,P^{M,\textup{ALG}}}[(L-t)_+]
\;=\;L_{\max}\,a_t
\;\ge\;
L_{\max}\,a_f^{-}\!\big(B;\,b_t\big).
\)
\end{corollary}

Corollary \ref{cor:hinge-lower} leads to the following CVaR lower bound.
\begin{corollary}[CVaR lower bound via the generalized Fano]
\label{cor:cvar-lower-bt}
Under the assumptions of Cor.~\ref{cor:hinge-lower}, by the Rockafellar--Uryasev identity \cite{rockafellar2000optimization}
$\mathrm{CVaR}_\alpha(L)=\min_{t\in\mathbb R}\{\,t+\tfrac{1}{1-\alpha}\mathbb{E}[(L-t)_+]\,\}$ and hence
\(
\;
\mathrm{CVaR}_\alpha(L)\ \ge\
\min_{t\in\mathbb R}\Big\{\,t+\frac{L_{\max}}{1-\alpha}\,
a_f^{-}\!\big(B;\,b_t\big)\Big\}.
\)

\emph{Proof.} Apply Cor.~\ref{cor:hinge-lower} for each $t$ and minimize over $t$.
\end{corollary}
Finally, Corollary \ref{cor:cvar-lower-bt} leads to a lower bound explicit in terms of mutual information via Pinsker's inequality.
\begin{corollary}[KL + mixture explicit CVaR lower bound]
\label{cor:cvar-lower-kl-mixture}
Let $f=\mathrm{KL}$ and choose the mixture reference $Q^\star=\mathbb{E}_{M\sim\mu}P^{M,\textup{ALG}}$.
Then $B=I_{\mu,\textup{ALG}}(M;X)$. By Pinsker’s inequality,
\(
D_{\mathrm{KL}}(\mathrm{Bern}(a)\|\mathrm{Bern}(b))\le B\ \Rightarrow\ |a-b|\le \sqrt{B/2}.
\)
Hence
\(
a_f^{-}(B;b)\ \ge\ [\,b-\sqrt{B/2}\,]_+.
\)
Plugging into Cor.~\ref{cor:cvar-lower-bt} gives
\[
\mathrm{CVaR}_\alpha(L)\ \ge\
\min_{t\in\mathbb R}\Big\{\,t+\frac{L_{\max}}{1-\alpha}\,
\big[b_t-\sqrt{\,I_{\mu,\textup{ALG}}(M;X)/2\,}\big]_+\Big\}.\]
\end{corollary}
Corollary~\ref{cor:cvar-lower-kl-mixture} formalizes that when the mutual information $I_{\mu,\textup{ALG}}(M;X)$ is small i.e., the data reveal little about M and the tail risk cannot be suppressed and $\mathrm{CVaR}_\alpha(L)$ admits a lower bound in terms of $I_{\mu,\textup{ALG}}(M;X)$.
\section{Relation to Prior Work}

Classical information-theoretic tools such as Le Cam, Assouad, and Fano provide general lower bounds for passive estimation \cite{tsybakov2009introduction}.
For interactive decision making, \cite{chen2024assouad} develops an interactive Fano framework that unifies several lower-bound paradigms (including relations to DEC-style bounds) under algorithm-dependent data laws.
Our results build on this line: relative to \cite{chen2024assouad}, we generalize their Fano method rather than using the expected risk or a single quantile event, we allow a broader class of bounded loss functionals and, in particular, deriving \(\textup{CVaR}\) lower bounds.
A complementary line studies tail criteria in non-interactive problems.
\cite{ma2024highprobability} prove risk-level-explicit minimax lower bounds for the \((1-\delta)\)-quantile of the loss,
\(
M(\delta)=\inf\Bigl\{r\in\mathbb{R}:\sup_{\theta\in\Theta}\mathbb{P}_\theta\bigl(L(\theta,\hat\theta)>r\bigr)\le \delta\Bigr\},
\)
which is equivalent to minimax \(\textup{VaR}_{1-\delta}\) since \(\mathbb{P}(L>r)\le \delta \iff \mathbb{P}(L\le r)\ge 1-\delta\) \cite{rockafellar2000optimization,rockafellar2002general}.
Moreover, \(\textup{CVaR}_{1-\delta}(L)\ge \textup{VaR}_{1-\delta}(L)\) for losses \cite{rockafellar2000optimization,rockafellar2002general}, so their minimax \(\textup{VaR}\) lower bounds imply minimax \(\textup{CVaR}\) lower bounds.
In contrast, our focus is on interactive lower bounds and on Bayesian \(\textup{CVaR}\) under the prior-predictive mixture induced by \((\mu,\textup{ALG})\), which is not a direct corollary of minimax quantile analyses.
The term 'Bayes' is also used in the risk-measure literature, but with a different meaning than our prior over instances.
\cite{embrechts2021bayes} studies when a risk measure can be written as \(\min_t \mathbb{E}[\ell(t,L)]\) for a suitable scoring loss \(\ell\), and show that Expected Shortfall, i.e.\ \textup{CVaR}, is essentially the only coherent example under a continuity condition; this does not address interactive data collection or \textup{ALG}-dependent information constraints.
Separately, \cite{thomas2019concentration,soma2020statistical,mhammedi2020pacbayes} give upper bounds for estimating/optimizing \textup{CVaR} from i.i.d.\ samples under fixed sampling models, whereas we derive information-theoretic lower bounds for the prior-predictive \textup{CVaR} induced by \((\mu,\textup{ALG})\) in interactive settings.

\section{Conclusion and Future Work}
We present an information-theoretic template for lower bounds on bounded loss functionals in ISDM under the prior–predictive law. It recovers interactive Fano’s tail/quantile bound and yields converse bounds for bounded transforms, including Bayesian \(\mathrm{CVaR}_\alpha\) with a KL/mutual-information specialization. Future work is to instantiate the template in bandits and episodic MDPs for closed-form Bayesian \(\mathrm{CVaR}\) limits, tighten constants using strong data processing or problem-specific contractions, and design algorithms matching these limits up to constants or logarithmic factors.

\IEEEpeerreviewmaketitle
 \newpage
\bibliographystyle{IEEEtran}

\bibliography{references}

\end{document}